# Utilizing a digital swarm intelligence platform to improve consensus among radiologists and exploring its applications.

Rutwik Shah MD, Bruno Astuto PhD, Tyler Gleason MD, Will Fletcher MD, Justin Banaga MD, Kevin Sweetwood MD, Allen Ye MD PhD, Rina Patel MD, Kevin McGill MD, Thomas Link MD PhD, Jason Crane PhD, Valentina Pedoia PhD, Sharmila Majumdar PhD

Center for Intelligent Imaging

Dept. of Radiology and Biomedical Imaging, UCSF

**ABSTRACT:**

**Introduction:** Radiologists today play a central role in making diagnostic decisions and labeling images for training and benchmarking Artificial Intelligence (A.I.) algorithms. A key concern is low inter-reader reliability (IRR) seen between experts when interpreting challenging cases. While teams-based decisions are known to outperform individual decisions, inter-personal biases often creep up in group interactions which limit non-dominant participants from expressing true opinions. To overcome the dual problems of low consensus and inter-personal bias, we explored a solution modeled on bee swarms.

**Methods:** Two separate cohorts; three board certified radiologists (Cohort 1), and five radiology residents (Cohort 2) collaborated on a digital swarm platform in real time and in a blinded fashion, grading meniscal lesions on knee MR exams. These consensus votes were benchmarked against clinical (arthroscopy) and radiological (senior-most radiologist) standards of reference using Cohen's kappa. The IRR of the consensus votes was then compared to the IRR of the majority and most confident votes of the two cohorts. IRR was also calculated for predictions from a meniscal lesion detecting A.I. algorithm.

**Results:** The attending cohort saw an improvement of 23% in IRR of swarm votes (k= 0.34) over majority vote (k=0.11). Similar improvement of 23% in IRR (k=0.25) in 3-resident swarm votes over majority vote (k=0.02), was observed. The 5-resident swarm had an even higher improvement of 30% in IRR (k=0.37) over majority vote (k=0.07). Swarm consensus votes also improved specificity by up to 50% compared to simple majority votes.

**Conclusion:** The swarm consensus votes outperformed individual and majority vote decision in both the radiologists and resident cohorts. The 5-resident swarm had higher IRR than 3-resident swarm indicating positive effect of increased swarm size. The attending and resident swarms also outperformed predictions from a state-of-the-art A.I. algorithm. Utilizing a digital swarm platform improved agreement among radiologists compared to standard of reference and can improve the quality A.I. training labels.

# Research in Context:

**Evidence before this study:**

There is a sizable body of work demonstrating that group-based decisions outperform individual opinions, in the fields of psychology and behavioral economics. We searched for similar evidence in prior medical literature in PubMed and Google Scholar using keywords "(Group OR teams OR crowds-based decisions) AND (medicine OR medical imaging)" to explore use of collective decisions within the clinical setting. Multiple research studies have reported superior performance of group decisions in both diagnostic (e.g., agreement in radiological or dermatological lesion identification) and interventional (e.g., prescribing antibiotics) tasks. However, most of these group assessments were performed post-hoc, aggregating opinions made by physicians in isolation and by using a collective metric such as an average or a majority vote. Other studies that allowed interactive group decisions, mentioned influence of inter-personal biases as a limitation.

**Added value of this study:**

We demonstrated that group consensus among expert participants using the novel swarm approach, outperformed their own individual assessments, when comparing to sets of reference standards. Moreover, this swarm consensus, obtained from real time interaction and in a bias free collaboration, outperformed other standard post-hoc metrics such as a majority vote. This observation was true for participants at two different levels of expertise (i.e., attendings and residents). Semantically, we also clarify the additional features of swarm intelligence over collective intelligence, terms which have been used inter-changeably in earlier studies. Additionally, we observed that the swarm also outperformed predictions made by a state-of-the-art A.I. model, trained on the same task.

**Implication of all available evidence:**

In combination with prior research, our work supports utilizing real time group-based decisions to improve consensus in the clinical setting, especially for interpretive tasks with low inter-rater reliability. The ability to express judgement-free opinion in interactive sessions is an important factor and merits further investigation. Finally, as A.I. becomes ubiquitous in medicine, the swarm approach allows obtaining diverse clinical opinions which can improve model performance and reduce algorithmic bias.

## INTRODUCTION:

Consensus amongst radiologists is key for accurate disease diagnosis, patient care and for avoiding inadvertent medical errors[1]. Guidelines from the National Academy of Medicine recommend a team-based diagnosis, considered superior to individual interpretation[2]. Obtaining high inter-rater reliability among experts can be challenging when interpreting complex multifactorial diseases and grading lesions on multiclass scales. The phenomenon of variable inter-rater reliability has been widely documented across imaging subspecialities [3-7], and can result in both missed diagnoses and limit appropriate medical intervention at the right time [8] (**Figure 1**).

Radiologists also perform an important role in training and benchmarking machine learning models. They classify and grade diseases, annotate lesions, and segmentation anatomical volumes on images [9,10]. Opinion of the radiologists is often considered as "ground truth" for training models and against which its performance is measured.

Given that annotation tasks can be time consuming, another approach is to have amateur labeling professionals (non-clinicians) annotate bulk of the images, with radiologists arbitrating discordant cases and performing a quality check of the dataset. However, use of non-experts is fraught with risks and can create noisy labels [11,12] or outright errors[13] which is consequential in high stakes artificial intelligence (A.I.) systems such as in medicine[14]. Numerous technical methods have been developed to mitigate effects of label noise. These include techniques for label cleaning and denoising [15,16], modifying loss functions [17,18], or data re-weighting [19-21]. However, none these methods fully mitigate the underlying cause of the noisy labels, which originate from interpersonal subjectivity at the time of label creation.

In both the approaches of expert and amateur data labeling, there is an assumption that the supervising radiologists being the experts, provide true value but this fails to factor in the disagreement observed between multiple experts themselves.

Some common methods used to decide the consensus answer in medicine include use of majority vote[22,23], most confident vote[24], arbitration[25], and the Delphi technique[26,27]. In this study we investigate a novel technique called swarm intelligence, to improve consensus among expert participants. Inspired from observations made in birds and insects[28-30], swarm intelligence is a method to find the optimal answer in a group of multiple autonomous agents, who collaborate in real time. This concept has found applications in fields ranging from economic forecasting[31], robotics[31] to imaging A.I.[32]

**Related work and key concepts:**

Collective intelligence or wisdom of the crowds, is defined as an emergent property of a quasi-independent multiagent system, where aggregated responses from the various agents outperforms individual responses[33]. This was perhaps best demonstrated by Galton's experiment demonstrating a crowd's average estimate of an ox's weight exceeding the best individual guess [34]. Multiple studies have demonstrated the phenomenon of collective intelligence and the various factors affecting it [35]. Individual conviction[36], level of expertise[37], cognitive diversity[38], personality traits[39] and social interaction[40] can all impact decision making in groups. We describe key concepts of team-based decision process in Table 1, relevant for understanding our study design.

Swarm intelligence (SI) is a specialized form of collective intelligence used to improve group decision-making in a wide range of biological species, from swarming bees and schooling fish to flocking birds. In recent years a technology called Artificial Swarm Intelligence (ASI) has been developed to enable similar benefits in networked human groups[41,42]. A software platform called Swarm was used in this study to enable networked human agents to make assessments by working together using the ASI technology. The software is designed to connect human agents with two distinguishing features; it requires *real-time participation* of all agents and it has a *closed loop feedback system* which updates and informs the agents of the combined group intent at each subsequent time step. It thus captures the dynamics of individual conviction, collaboration, negotiation, and opinion switching and is not simply a post-hoc majority or an average vote analysis.

The primary aim of our study was to examine the effect of *synchronous*, blinded *nonsocial* interaction among clinical *experts* at different levels of expertise (radiologists, radiology residents), on a *specific task* (evaluation of meniscal lesion on knee MR) while answering a *fixed questionnaire,* and measure its effect on inter-rater reliability. Our secondary aim was to examine the effect of the number of participants (swarm size) in improving inter-rater reliability.

**METHODS:**

**Radiographic and Clinical Dataset:**
The present study was conducted using previously acquired knee MRIs, and corresponding clinical notes of 36 subjects enrolled for a longitudinal research study(Arthritis Foundation- ACL Consortium)[43]. Subjects were recruited and scanned at one of the three sites: University of California, San Francisco (San Francisco, CA), Mayo Clinic (Rochester, MN), and Hospital for Special Surgery (New York, NY). All subjects underwent arthroscopic evaluation and repair of the affected knee by an orthopedic surgeon, who recorded findings in the various compartments (meniscus, cartilage, bone, and ligaments) for lesions.

Distributions of patient demographics were age=42.79±14.75 years, BMI=24.28±3.22Kg/m2, 64%/36% male/female. Study subjects were recruited with age>18 years and exclusion criteria being- concurrent use of any investigational drug, fracture or surgical intervention in the study knee, and any contraindications to MR. All subjects signed written informed consent approved by the Committee on Human Research of the home institution. The study was approved by the Institution Review Board.

**Study participants (radiologists and radiology residents) and task:**

Two cohorts of readers were recruited to evaluate the knee scans at multiple timepoints (**Figure 2**). All readers examined only the sagittal CUBE sequence on the institutional Picture Archiving and Communication System (PACS). They were asked to answer the same question for each exam; "Select the regions of the meniscus where a lesion is observed", where a lesion was defined as Whole Organ Magnetic Resonance Imaging Score (WORMS) >0[44]. The six possible answer choices given were 1) none, 2-5) any one of the four meniscal horns (anterior and posterior; medial and lateral horns) compartments or 6) more than one compartment.

Cohort 1 included 3 board certified musculoskeletal radiology attendings (averaging 19 of experience, range 4-28) who read the MRI scans at two timepoints. First, at baseline, they independently graded the scan individually, also giving a self-reported confidence score for their reads (scale: 1 to 10). After a 15-day washout period, all 36 exams were reassessed by the attendings, while participating simultaneously in a swarm session (Unanimous AI, San Francisco), in real time.

Cohort 2 included 5 radiology residents (PG Year 3-5). Similar to the attendings, they too first graded the scans independently at baseline with self-reported confidence scores. After a 15-day washout period, all 36 scans were reassessed by all 5 residents for a second time while participating simultaneously in a swarm session. After another 15-day washout period, 3 of the 5 residents (partial Cohort 2) reassessed the 36 scans for a third time while participating in a second swarm session. This was done to measure the effect of swarm size on the inter-rater reliability.

**Swarm platform:**

To obtain the consensus answer of our participating radiologists and trainees, we utilized Swarm platform (Unanimous AI, San Francisco), a platform which is modeled on the decision-making process of honeybees[45].

The platform allows multiple remotely located participants to collaborate in a blinded fashion over the internet, in real time.

The platform consists of 2 key components: 1) a web-based application and 2) a cloud-based server that runs the proprietary swarm algorithm. Participants log into an online swarm session, using a standard web browser and answer questions on the platform's hexagonal graphical user interface (GUI). The GUI captures real-time inputs from the full set of participants and provides immediate feedback based on the output generated from the swarm algorithm, essentially creating a closed-loop feedback system (**Figure 3**).

Using this system, both the cohorts answered questions in real time by collaboratively moving a graphical puck to select among a set of answer options. Each participant provided input by moving a graphical magnet to pull on the puck, thereby imparting their personal intent on the collective system. The preference is recorded as a continuous stream of inputs rather than just as a static vote. The conviction of each individual participant is indicated by distance between their magnets and the puck (strong versus weak pull). The net pull of all participants on the moves the puck in that direction, until a consensus is reached on one of the answer choices. The output of the collective answer is therefore also updated on the GUI in real time, as observed by the changing trajectory of the puck during an active swarm session. Because all users adjust their intent continuously in real time, the puck moves based on the complex interactions among all participants, empowering the group converge in synchrony.

Meanwhile, the swarm algorithm evaluates each user's intent at each instant by tracking the direction and strength of the pull of their magnets while comparing it with other participants. This is then used to i) compute the consensus answer at each time step based on collective preferences and ii) to provide instantaneous feedback to participants in the form of an updated puck trajectory, allowing them to stay with or switch their original answer choice, given the evolving group decision. The consensus decision computed by the swarm algorithm considers various factors such as i) the number of swarm participants ii) the participants' initial preferences iii) participants' behavior (consistent versus changing in opinion) iv) level of conviction and v) type of answer choices (ordinal versus categorical).

**Swarm sessions:**

Cohort 1 (3 MSK radiologists) participated in a single swarm session, after a wash out period after the individual assessment of the knee scans. Cohort 2 (radiology residents) participated in two consecutive swarm sessions post a washout period after their individual assessment. The first resident swarm session had 5 residents. The second resident swarm session had 3 residents and was conducted to measure the effect of the swarm size.

To answer each question during our study, all participants in both cohorts were allowed 60 seconds to first review the knee scan and then another 60 seconds to actively participate in the swarm session, collaborate and provide their consensus answer. In some instances of strong opposing opinions, a swarm may not be able to reach an answer within the time allotted to decide, in which case the platform records it as a "no consensus". All the participants in both the cohorts were blinded to each other and didn't communicate during the session to prevent any form of bias.

**A.I. Model inference:**

To benchmark a state-of-the-art AI model against swarm performance of the radiologists and residents, we ran the model over the same set of 36 knee MR scans (sagittal CUBE sequences only). An A.I. pipeline for localization and classification of meniscus lesions was trained and validated on a retrospective study conducted on 1435 knee MRIs ($n = 294$ patients, mean age, $43 \pm 15$ years, 153 women)[46]. The AI pipeline consisted of a V-Net convolutional deep learning architecture to generate segmentation masks for all four meniscus horns, that were used to crop smaller sub-volumes containing these regions of interest (ROIs). Such sub-volumes were used as input to train and evaluate three-dimensional convolutional neural networks

(3DCNNs) developed to classify meniscus abnormalities. Evaluation on the holdout set yielded sensitivity and specificity of 85% and 85% respectively on a binary assessment ("Lesion" or "No Lesion").

**Statistical analysis:**

All responses were binned into 3 classes (none, one compartment, more than 1 compartment) to enable comparisons between individual participant votes, swarm votes and A.I. predictions. Confidence scores of the individual responses among participants of the same cohort, were harmonized to evaluate for internal consistency using Cronbach's alpha. Sensitivity, specificity and Youden index (measure of accuracy) was calculated for presence or absence of lesions.

The first time point responses were then used to calculate the majority vote and choose the most confident voter in each cohort. Cohen's kappa (k) values were tabulated with mean, standard deviation, and confidence intervals, bootstrapped 100 times re-sampling a full set of cases from the observations, to evaluate inter-rater reliability as described below.

1. <u>Attending inter-rater reliability compared with clinical standard of reference (IRRc)</u> – The first set of analyses was conducted comparing attending (Cohort 1) responses to arthroscopic notes considered as clinical standard of reference (SOR). IRR of the individual attendings, their majority vote and the most confident vote was calculated. The IRR of the attending swarm vote was also computed with respect to clinical SOR as well.

2. <u>Resident inter-rater reliability compared with clinical standard of reference (IRRc)-</u> The second set of analyses was conducted comparing residents (Cohort 2) to the clinical SOR. Inter-rater reliability of the individual residents, their majority vote and the most confident vote was calculated. The IRR of the swarm vote was also computed with respect to clinical SOR for both the 5 resident and 3 resident swarm votes.

3. <u>Resident inter-rater reliability compared with radiological standard of reference (IRRr)-</u> In many cases, clinical ground truth from surgical evaluation of lesions may not be available. Additionally, there may be low inter-rater reliability between radiologists and surgeons as well. In such instances, the interpretation of an experienced radiologist is often considered as standard of reference, especially when evaluating trainees.

To evaluate for swarm performance in such scenarios, we considered the responses of our senior most participating attending as radiological standard of reference. IRR of the individual residents, their majority vote and the most confident vote was calculated. The IRR of the swarm vote was also compared with radiological SOR for both the 5 resident and 3 resident swarm votes

4. <u>Comparing A.I. predictions with clinical and radiological standards of reference (IRRc and IRRr)</u>- Similar to the resident and attending cohorts, the predictions of the model inference were compared with both the clinical and radiological SOR.

**RESULTS:**

The class balance as per clinical standard of reference was as follows: normal (15/36 exams), lesion in one compartment (13/36 exams), lesions in more than compartment (8/36 exams). The class balance as per radiological standard of reference was as follows: normal (8/36 exams), lesion in one compartment (8/36 exams), lesions in more than compartment (20/36 exams).

Both the attending and resident cohorts show excellent internal consistency with Cronbach's alpha of 0.91 and 0.92 respectively. The sensitivity, specificity and Youden index are described in **Table 2**. Both the cohorts had high sensitivity in detecting meniscal lesions, comparable between the majority votes, most confident votes, and the swarm votes. The swarm votes showed an improvement in specificity in all scenarios and an increase in specificity was also observed with an increase in the resident swarm size. The attending swarm votes saw specificity improve by 40% (53.3%) over the attending majority vote (13.3%). The 3 resident swarm demonstrated an improvement in specificity of 20% over the majority vote and the most confident vote, for comparisons against the clinical SOR. The 3 resident swarm also showed an improvement in specificity of 37.5% over the majority vote and most confident votes, for comparisons based on radiological SOR. Similarly, the 5-resident swarm vote showed a specificity of 33% (based on clinical SOR) and 50% (based on radiological SOR), much higher than either the 5-resident majority and most confident vote. This has important clinical implications in preventing overdiagnosis of lesions.

Bootstrapped Cohen's kappa of the attending and resident cohorts' inter-rater reliability with the clinical and radiological standard of reference are mentioned in **Table 3,** with corresponding 95% confidence intervals. The swarm consensus votes consistently showed higher IRR than the individual voters, their majority vote, and the most confident voter. Superior IRR of swarm votes was observed for both the attending and resident cohorts. More importantly, an increase in swarm IRR was seen in both $IRR_c$ and $IRR_r$. The swarm methodology thus improved agreement with either standard of reference, indicating its usefuless for assessment, even in scenarios when clinical and radiological observations may have discordance. An increase in IRR was also observed with an increase in resident swarm size. Interestingly, the 5-resident swarm $IRR_c$ agreement was at a comparable level to the 3-attending swarm $IRR_c$. While the absolute kappa values reported in this study are in the slight to fair range, these should be viewed in light of the limited imaging exam (single sagittal MR sequence only) which was made available for the participants.

The IRRc for individual attendings ranged from k=0.08 to 0.29. The 3 attending swarm vote IRRc was higher compared to the 3 attending majority vote and the 3 attending most confident vote (**Figure 4**). Agreement on detecting normal cases increases significantly from 13% for majority vote (2/15) to 53% (8/15) for swarm vote. Since the senior most radiologist was part of this cohort, no IRRr was calculated for the attendings.

IRRc for individual resident responses ranged from k=0.01 to 0.19 and was lower compared to the attendings. The 3 resident-swarm vote IRRc was higher compared to the 3 resident majority vote and 3 resident most confident vote (**Figure 5**). The majority vote and most confident vote failed to identify any normal cases. Agreement on detecting normal cases is 20% (3/15) for swarm vote. The 5 resident-swarm vote IRRc was again higher than the 5 resident majority vote and the 5 resident most confident vote. The 3 resident majority vote and most confident vote failed to identify any normal cases. The 5 resident- majority vote failed to identify any normal cases. Agreement on detecting normal cases increases by 33% (5/15) for swarm vote.

IRRr for individual resident responses vs radiological observation ranged from k=0.09 to 0.22. This was higher compared to the resident IRRc, indicating they had better agreement with their direct trainers i.e., the radiology attendings than with the orthopedic surgeons. In line with the earlier findings, both the 3 and 5 resident swarm vote IRRr was higher than their respective majority votes and most confident votes. (**Figure 6**). The majority vote and most confident vote failed to identify any normal cases. Agreement on detecting normal cases was 37.5% (3/8) for swarm vote. The majority vote and most confident vote failed to identify any normal cases. The 5 resident- majority vote failed to identify any normal cases. Agreement on detecting normal cases increased for the swarm vote in both size cohorts.

As opposed to the 3 resident and 3 attending swarms, the 5 resident swarm failed to reach a consensus in one exam, in the allotted time. This single occurrence was not enough to conclusively comment on relationship of swarm size and optimal time for decision and was subsequently excluded during comparisons with the majority and most confident votes.

A.I. predictions from the model inference had an IRRc of k= 010 and IRRr of k=0.15. This is comparable to the range of individual resident inter-rater reliability.

**DISCUSSION:**

Multiple studies have reported varying IRR among radiologists in interpreting meniscal lesions[47]. Differences in opinions can occur based on location, zone, tissue quality and severity of lesion. Shah et al. reported prevalence and bias adjusted kappa ranging from poor for medial meniscus zone 1(k= 0.22) to excellent for lateral meniscus zone 3 (k= 0.88)[48]. Some imaging related factors for the low agreement include limited image resolution, motion artifacts and the limited time afforded to radiologists for image interpretation under an ever increasing workload[49].

Arthroscopic evaluation is often considered as the clinical standard of reference for evaluating radiological reads[50]. However, surgeons have a narrower field of view during arthroscopy and lack the ability to view region of interest in multiple orientations (sagittal, axial, coronal) simultaneously. These factors limit consideration of surgical observations as reliable clinical ground truth.

Additionally, there may be a lag time of days to weeks between imaging and arthroscopy allowing improvement or deterioration of lesion, and which can further limit agreement with their radiology colleagues. Kim et al. reported inter-method reliability (radiology- arthroscopy) kappa values ranging from 0.40 to 0.71 depending on laterality of lesion and presence of ACL tears[51]. Such differences in opinions are problematic for generating clinical consensus and defining ground truth labels for A.I. training. Given that the radiologist's report and arthroscopy evaluations can have some disagreement, we examined use of swarm methodology against a radiological standard of reference (senior-most radiologist) as well.

Multiple investigators in the past have advocated use of consensus voting to improve medical diagnoses [52] and demonstrated superior performance of majority or average vote [53]. However, no study till date had compared consensus votes from a real time blinded collaboration to a post-hoc majority vote. There have been varying opinions on what exactly improves the accuracy in a crowds-based answer; the effect of social interaction[54] or pure statistical aggregation. Social interaction can be further complicated by the interpersonal biases which can either improve or worsen crowd performance[55,56]. Thus, it is pertinent to understand the exact influence of these factors especially when they are applied to make clinical decisions.

Our current study explored these questions by first performing non-social interactions between blinded participants at equal levels of expertise (radiologists or residents' cohorts), in a bias free environment. Next the resident cohort repeated a swarm session with fewer participants, to measure the effect of group size on the responses. Our results show both the group size and interaction influence performance, although conducting negotiations for the optimal answer under *anonymization*, was key for resisting peer pressure.

A key aspect of our study was to evaluate performance of an A.I. model on the same set of 36 knee exams. This model had been trained and tested on labels created by multiple radiologists and residents at our institution over time. The A.I. $IRR_c$ was k=0.10 and was comparable to the $IRR_c$ of the 3-residents most confident vote. The A.I. $IRR_r$ was k=0.15, comparable to the $IRR_r$ of the 3-resident most confident vote. In other words, the AI performance is already as good as its trainers. In both cases however, the kappa was significantly lower than the kappa of either the resident or the attending swarms. A useful strategy to improve model performance beyond its current results, would be to use swarm votes as labels in the training datasets. Leveraging swarm intelligence for A.I. training would provide higher quality labels which are more accurate, mitigate the problem of noisy labels, and reduce the need for large training datasets as currently needed for most deep learning models.

Swarm voting improved IRR by up to 32% in our study, which was based on a specific imaging modality (MR), and for a specific task of evaluating meniscal lesions. It would be important to investigate the increase in diagnostic yield by real time consensus voting, in other diagnostic imaging scenarios across different modalities as well. The swarm platform would be a useful tool for expert radiologists to collaborate and evaluate complex or ambiguous cases. A potential first application would be for imaging workflows where multiple reads are already mandated, such as for double reads for breast mammograms, as practiced in Europe[57].

Our study had a few limitations. While we aimed to simulate the regular radiology workflow with use of PACS, it did not capture the entire experience given the time constraints to run the swarm sessions. Normally, radiologists have access to multiple sequences and views in an MR exam, with prior exams and other relevant clinical notes for comparison. We speculate the inter-rater reliability in our study would have been higher and in line with other reported studies, with availability of complete MRI exams.

Given scheduling challenges in the pandemic, we performed only necessary swarm sessions as required for this pilot study. While we observed improvements in swarm agreement with both the standards of reference, the overall dataset in this study was not large enough to power a statistically significant difference over individual or majority votes.

Though we were able to observe improved inter-rater reliability and specificity with increase in swarm size (five versus three resident swarm), further investigation with additional participants is warranted to estimate optimal group size. Given the limited availability of expert radiologists, it will be important to understand if diagnostic gains made with larger groups peak at a certain participant number.

In conclusion, utilizing a digital swarm platform improved consensus among radiologists and allowed participants to express judgement free intent. This novel approach outperforms traditional consensus methodologies such as a majority vote. Future direction of our work includes performing serial swarm sessions to generate more accurate labels for A.I. training. We also aim to explore the swarm platform for evaluating trainee performance. Residents at our center can self-assess their diagnostic opinions with peers, and the training program can assess group performance across cohorts over time, in an objective manner.

**Acknowledgement:** R33 AR073552, grant from the NIH was utilized for conducting this study. We would like to thank Unanimous AI, for providing us pro-bono access to the Swarm software platform for our study.

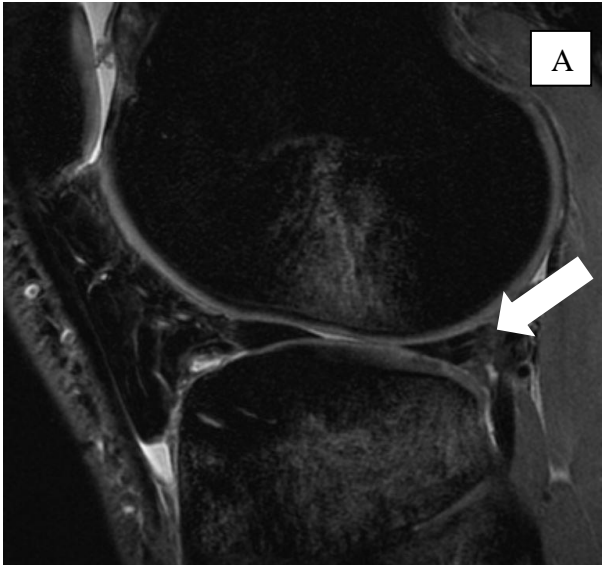 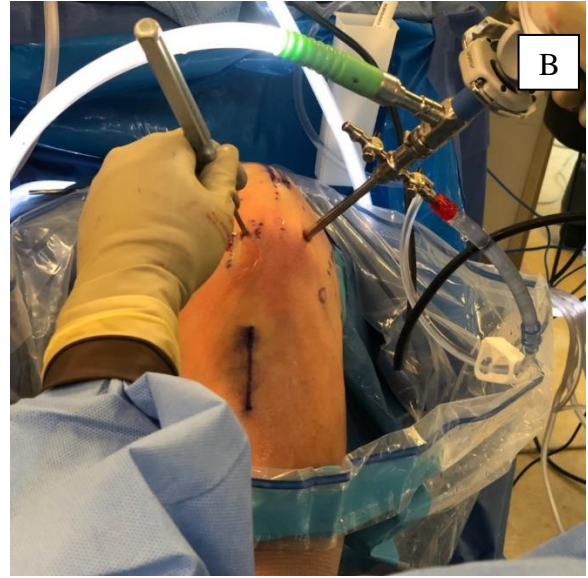

**Figure 1:** A) Sagittal sequence of a knee MR exam evaluated by multiple subspeciality trained musculoskeletal radiologists (arrow pointing to ambiguous meniscal lesion) had discordant impressions of the presence and grade of lesions. B) Swarm platform was used to derive consensus for location of lesions, which matched with the arthroscopic findings considered as a standard of reference.

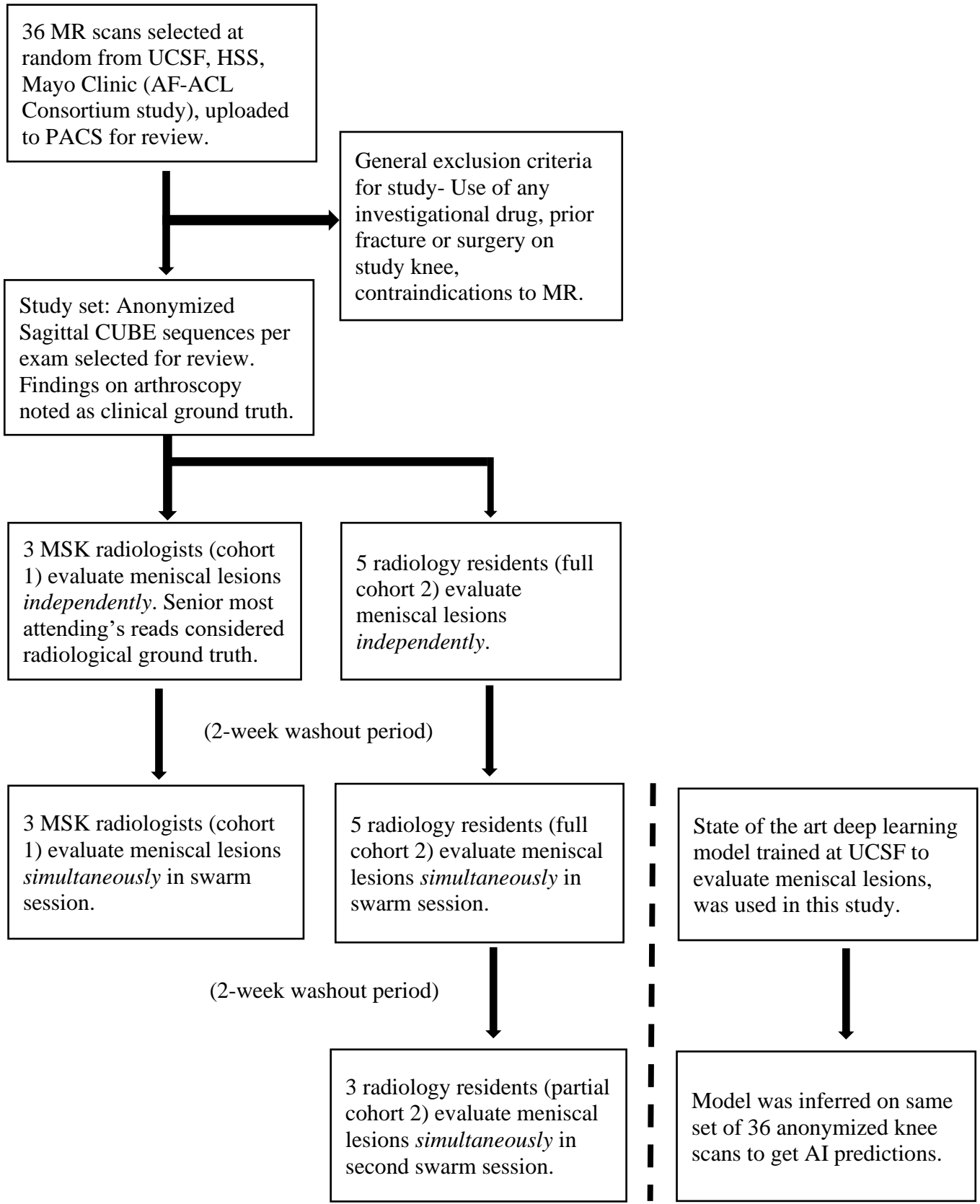

**Figure 2:** Flowchart of various steps in the study. 36 anonymized knee scans (sagittal CUBE sequences) were reviewed by a cohort of three MSK trained radiologists and another cohort of five radiology residents, independently at first and then in swarm sessions. A deep learning model trained to evaluate meniscal lesions also inferred the same of 36 knee scans to obtain AI predictions for comparison.

| Key Concepts | Options in teams-based decision making |
|---|---|
| Time of participation | Agents can participate in the prescribed activity asynchronously and then have results calculated post-hoc e.g., majority vote or average vote tabulation. Or, agents can participate synchronously, where all participants answer questions at the same time, without exception. This is a key feature of the digital swarm platform. |
| Expertise | Participating agents can all be domain experts (e.g., radiologists, radiology residents trained in specialized image interpretation), or non-experts who may not possess specialized expertise relevant to the task at hand. |
| Scope of task | The scope of the task for answering each question can be broad including multiple tasks (review images, clinical notes, and lab reports) or narrow and include a single task (image review) only. |
| Questionnaire | The set of questions asked to the agents can be fixed and consistent for each item. or can be adaptive based on previous responses, as seen in the Delphi technique. |
| Communication | Communication can be either social or non-social. Social interaction allows agents to assess other's interests, preferences, and also influence each other while performing the task at hand. This can lead to various inter-personal biases which can negatively impact overall results[55]. In contrast, non-social interaction allows agents to know group intent while being blinded to the identity, preferences, and level of expertise of other participants. |

**Table 1:** Key concepts in teams-based decision making. Swarm intelligence requires real-time collaboration of all participants, with constant feedback of the group intent. Our study was designed to also be asocial to prevent any inter-personal bias.

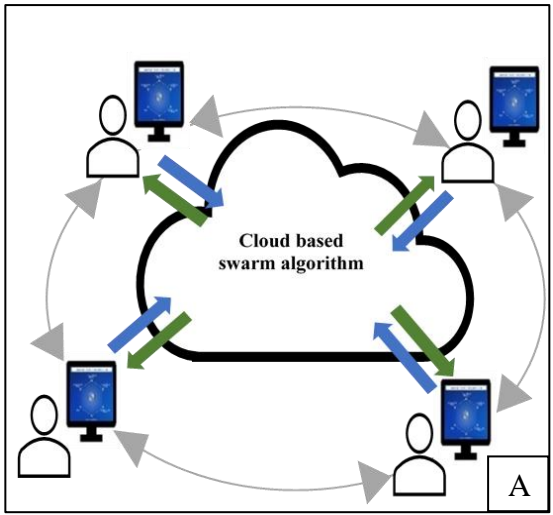 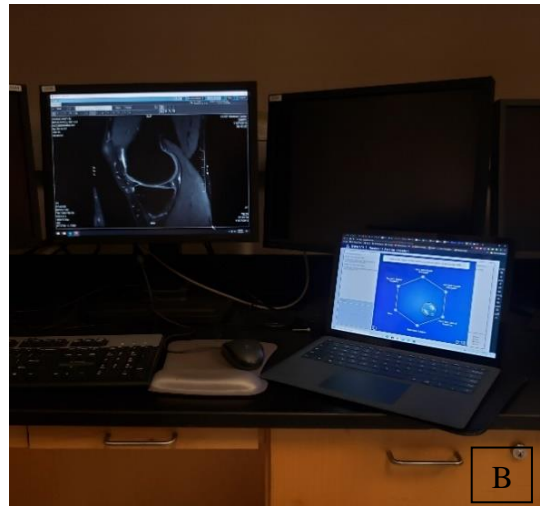
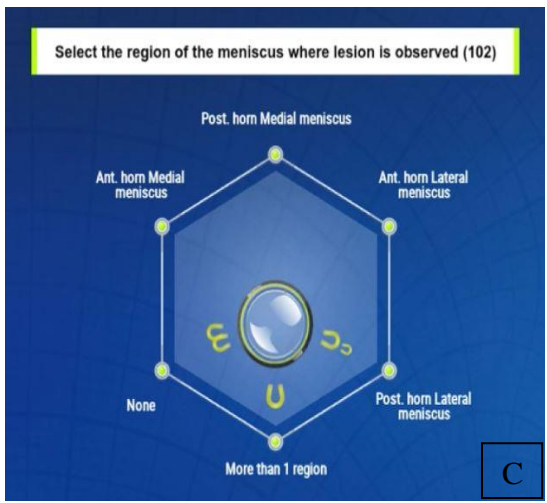 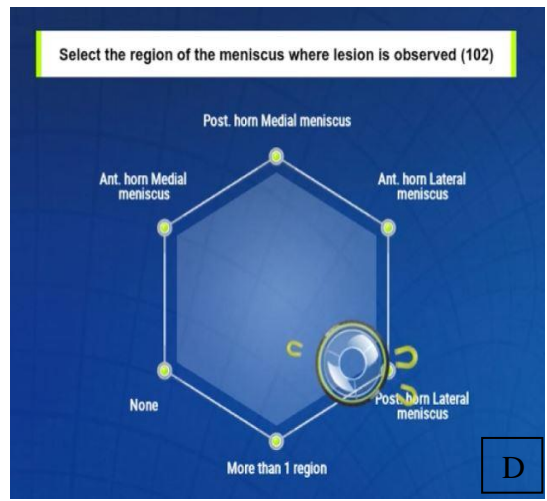

**Figure 3:** A) Schematic of the swarm platform. Multiple remote users are connected to each other in real time, via the web application. Inputs from users (blue arrows) are sent to the cloud server which runs the swarm algorithm, which then sends back continuous a stream of output (green arrows) to users in a closed loop system. B) Setup of the swarm session- Participants accessed the knee exams on a PACS workstation and logged into swarm sessions via a separate device. C) Early time point in a session- multiple users pulling central puck in opposing directions using virtual magnets as seen in the graphical interface. D) Late time point in same session- users then converge onto a single answer choice after some negotiation and opinion switch.

|  | Clinical Standard of Reference | | | Radiological Standard of Reference | | |
| --- | --- | --- | --- | --- | --- | --- |
|  | Sensitivity | Specificity | Youden index | Sensitivity | Specificity | Youden index |
| 3 attending majority vote | 100% | 13.3% | 0.13 | N/A | N/A | N/A |
| 3 attending most confident vote | 95.2% | 33.3% | 0.28 | N/A | N/A | N/A |
| 3 attending swarm vote | 90.4% | 53.3% | 0.43 | N/A | N/A | N/A |
| 3 resident majority vote | 100% | 0 | 0 | 100% | 0 | 0 |
| 3 resident most confident vote | 100% | 0 | 0 | 100% | 0 | 0 |
| 3 resident swarm vote | 100% | 20% | 0.20 | 100% | 37.5% | 0.37 |
| 5 resident majority vote | 100% | 0 | 0 | 100% | 0 | 0 |
| 5 resident most confident vote | 95% | 6.6% | 0.01 | 96.2% | 12.5% | 0.08 |
| 5 resident swarm vote | 95% | 33% | 028 | 92.5% | 50% | 0.42 |
| AI prediction | 100% | 13.3% | 0.13 | 100% | 25% | 0.25 |

**Table 2:** Sensitivity, specificity and Youden's index for binary outputs for the attending and resident cohorts. Swarm consensus votes had higher specificity than majority vote or most confident vote for both cohorts in all scenarios. The 5 resident swarm also shows higher specificity than the 3 resident swarm vote.

|  | Mean (std) Kappa | 95% CI | p-value |
|---|---|---|---|
| **3 Attending Majority Vote versus Clinical SOR** | 0.11 (0.06) | [0.02 - 0.24] | 0.05 |
| **3 Attending Most Confident Vote versus Clinical SOR** | 0.19 (0.09) | [0.02 - 0.35] | 0.02 |
| **3 Attendings Swarm versus Clinical SOR** | 0.34 (0.11) | [0.16 - 0.53] | 0.18 |
| **3 Residents Majority Voting versus Clinical SOR** | 0.02 (0.04) | [-0.07 - 0.09] | 0.16 |
| **3 Resident Most Confident Vote versus Clinical SOR** | 0.08 (0.04) | [0.02 - 0.17] | 0.85 |
| **3 Resident Swarm versus Clinical SOR** | 0.25 (0.09) | [0.08 - 0.47] | 0.85 |
| **5 Residents Majority Vote versus Clinical SOR** | 0.07 (0.06) | [-0.04 - 0.19] | 0.79 |
| **5 Resident Most Confident Vote versus Clinical SOR** | 0.12 (0.07) | [-0.02 - 0.26] | 0.72 |
| **5 Residents Swarm versus Clinical SOR** | 0.37 (0.10) | [0.16 - 0.61] | 0.54 |
| **3 Resident Majority Vote versus Radiological SOR** | 0.27 (0.10) | [0.09 - 0.49] | 0.24 |
| **3 Resident Most Confident Vote versus Radiological SOR** | 0.15 (0.10) | [-0.03 - 0.37] | 0.41 |
| **3 Resident Swarm versus Radiological SOR** | 0.36 (0.14) | [0.08 - 0.63] | 0.09 |
| **5 Resident Majority Vote versus Radiological SOR** | 0.32 (0.09) | [0.16 - 0.52] | 0.74 |
| **5 Resident Most Confident Vote versus Radiological SOR** | 0.14 (0.14) | [-0.11 - 0.35] | 0.18 |
| **5 Resident Swarm versus Radiological SOR** | 0.39 (0.12) | [0.15 - 0.63] | 0.03 |
| **AI versus Clinical SOR** | 0.10 (0.09) | [-0.11 - 0.28] | 0.001 |
| **AI versus Radiological SOR** | 0.15 (0.14) | [-0.13 - 0.45] | 0.015 |

**Table 3:** Cohen's kappa values in various comparisons of attendings, residents and AI versus clinical and radiological standards of reference (SOR). For both the attendings and residents the swarm consensus vote has better agreement than either the majority vote or most confident voter.

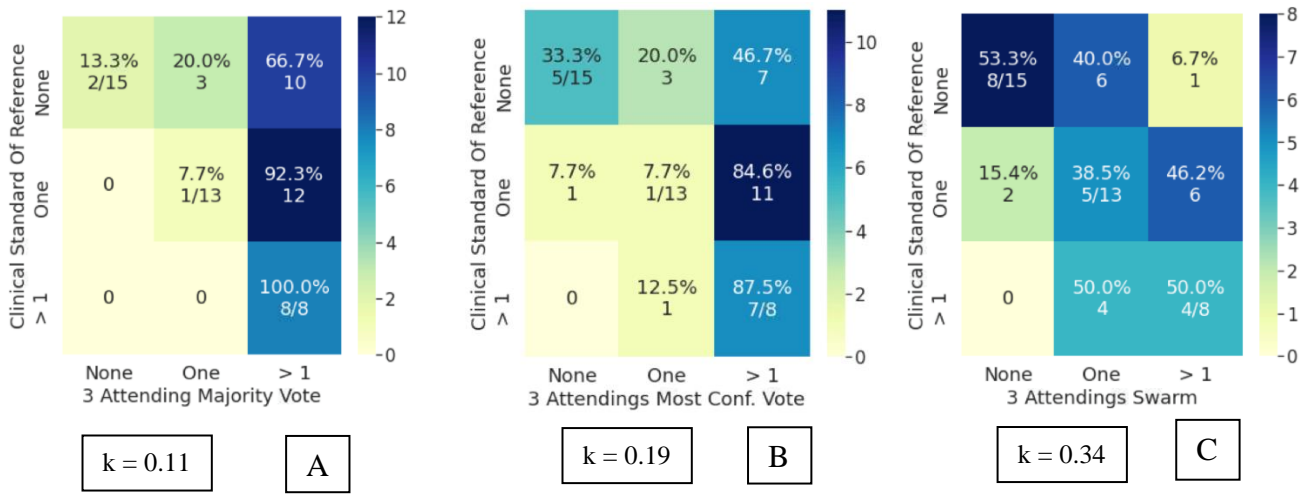

**Figure 4:** Attendings grading compared to clinical standard of reference. A) Confusion matrix (CM) for 3 attending majority vote (kappa:0.12). B) CM for 3 attending most confident vote (kappa:021). C) CM for 3 attending swarm vote (kappa:0.35).

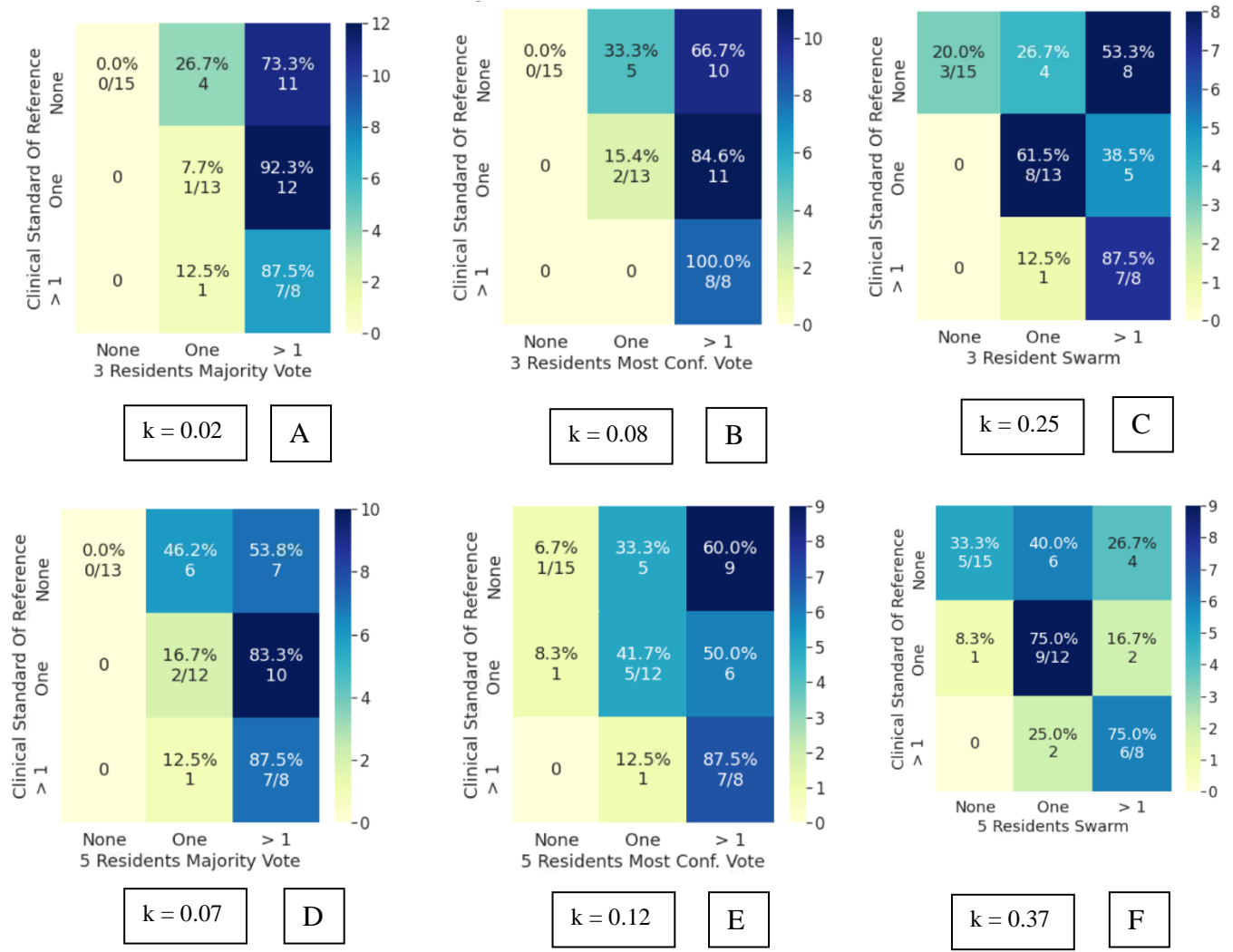

**Figure 5:** Residents grading compared to clinical standard of reference. A) Confusion matrix (CM) for 3 resident majority vote (kappa: 0.01) B) CM for 3 resident most confident vote (0.07). C) CM for 3 resident swarm vote (kappa: 0.24) D) CM for 5 resident majority vote (kappa: 0.05) E) CM for 5 resident most confident vote (0.12). F) CM for 5 resident swarm vote (kappa: 0.37).
Note: The 5 resident swarm was unable to obtain a consensus in one exam. This exam was excluded during inter-rater reliability comparisons of 5 resident majority vote and 5 resident most confident vote for parity.

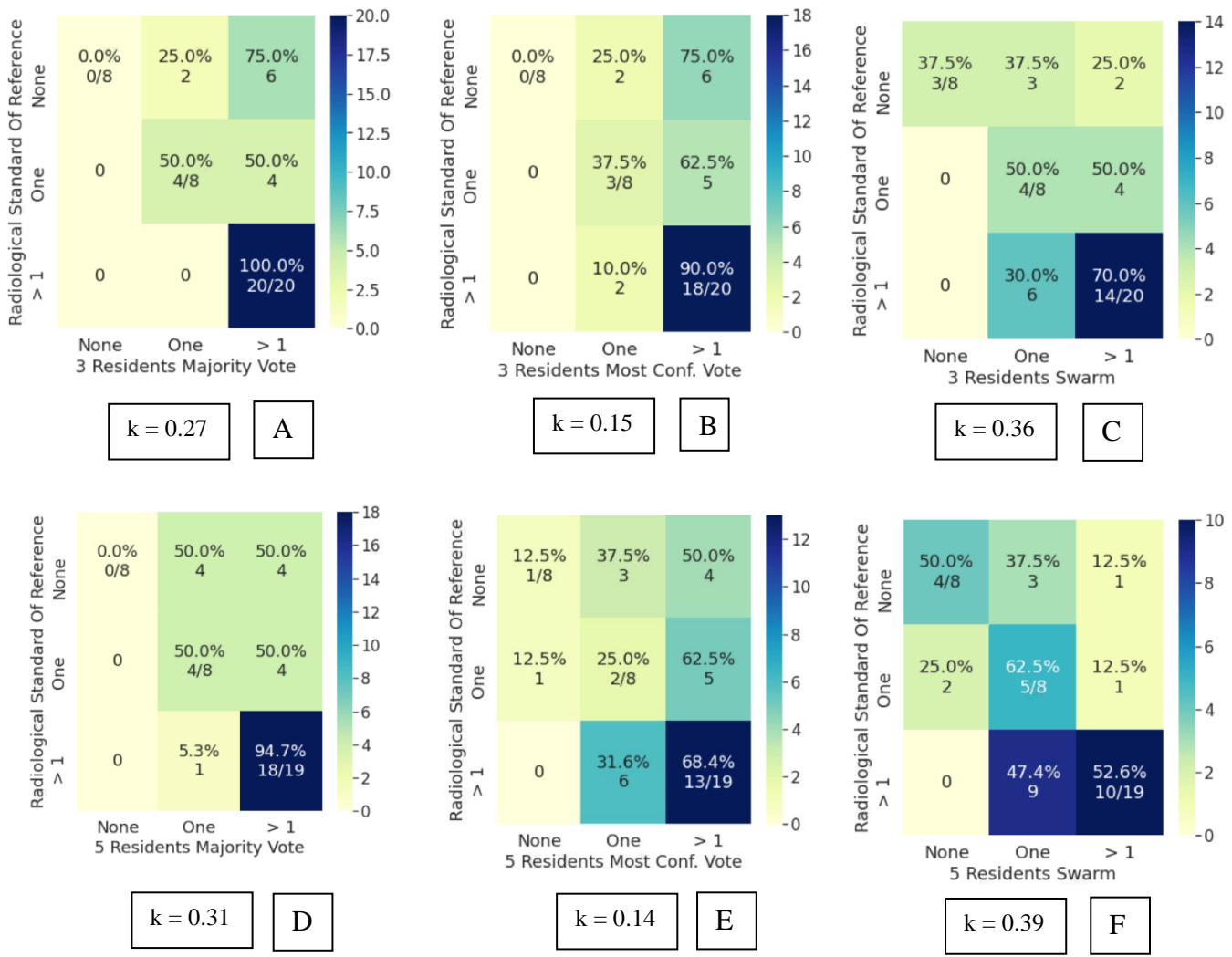

**Figure 6:** Residents responses compared to radiological standard of reference. A) Confusion matrix (CM) for 3 resident majority vote (kappa: 0.27) B) CM for 3 resident most confident vote (0.2). C) CM for 3 resident swarm vote (kappa: 0.38) D) CM for 5 resident majority vote (kappa: 0.31) E) CM for 5 resident most confident vote (0.12). F) CM for 5 resident swarm vote (kappa: 0.42).

Note: The 5 resident swarm was unable to obtain a consensus in one exam. This exam was excluded during inter-rater reliability comparisons of 5 resident majority vote and 5 resident most confident vote for parity.

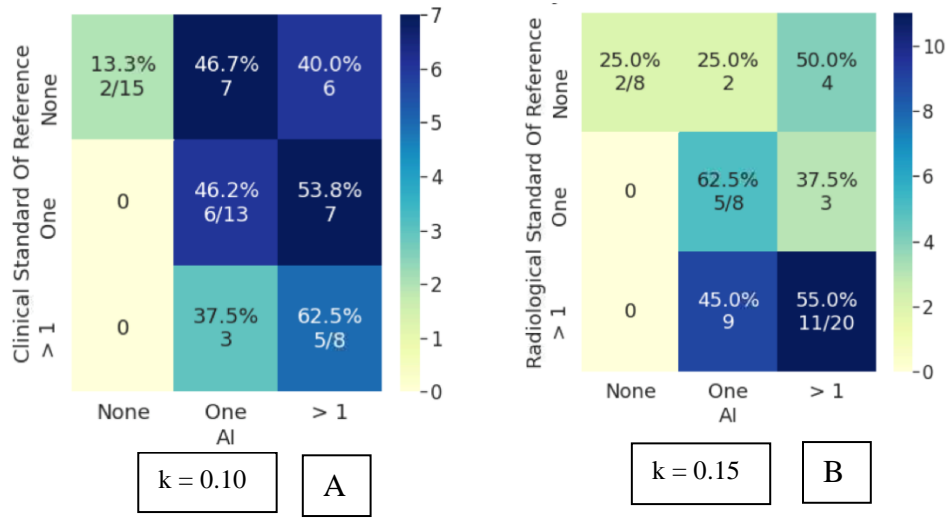

**Figure 7:** AI predictions comparisons. A) Confusion matrix for AI predictions compared to clinical standard of reference (kappa:0.15). B) Confusion matrix for AI predictions compared to radiological standard of reference (kappa: 0.17). Swarm votes of residents outperform AI in both sets of comparisons.